\documentclass[12pt]{article}
\usepackage{graphicx,color}
\usepackage{amssymb}
\usepackage{amsmath}

\usepackage{times}
\usepackage[T1]{fontenc}
\normalfont
\usepackage{mathptm}
\usepackage{fourier} 

\newcommand{\MeV}{\mbox{~MeV}}

\begin{document}
\baselineskip 0.7cm
%
\begin{titlepage}
\begin{center}

\begin{flushright}
SU-HET-05-2016\\
\end{flushright}

\vskip 2cm

{\Large \bf 
On neutrinoless double beta decay in the $\nu$MSM}

\vskip 1.2cm

{\large 
Takehiko Asaka$^1$,
Shintaro Eijima$^2$, and Hiroyuki Ishida$^3$
}

\vskip 0.4cm

$^1${\em
  Department of Physics, Niigata University, Niigata 950-2181, Japan
}

$^2${\em
  Institute of Physics, \'Ecole Polytechnique
F\'ed\'erale de Lausanne, CH-1015 Lausanne, Switzerland
}

$^3${\em
  Graduate School of Science and Engineering, Shimane University, 
Matsue 690-8504, Japan
}

\vskip 0.2cm

(June 22, 2016)

\vskip 2cm

\vskip .5in
\begin{abstract}
    We consider the neutrinoless double beta ($0\nu \beta \beta$) decay 
    in the $\nu$MSM, in which three right-handed
    neutrinos with masses below the electroweak scale are additionally 
    introduced to the Standard Model.
    In this model there appear
    three heavy neutral leptons $N_1$, $N_2$, and $N_3$ corresponding
    to right-handed neutrinos.  It has been known that the lightest
    one $N_1$ with keV mass, which is a candidate for dark matter,
    gives a negligible contribution to the $0 \nu \beta \beta$ decay. 
    By contrast, the heavier 
    ones $N_2$ and $N_3$, which are responsible to the
    seesaw mechanism of neutrino masses and baryogenesis, give the destructive contribution 
    (compared with one from active neutrinos). This is because their mass degeneracy at high precision
    has been assumed, which is expected by analytical studies of baryogengesis.
    In this analysis, we find that
    the effective mass of the $0\nu \beta \beta$ decay 
    becomes larger than one from active neutrinos
    due to the $N_2$ and $N_3$ constructive contribution
    when the mass difference becomes larger
    and the mass ordering of active neutrinos is inverted.
    Such a possibility will be explored 
    by the current and near future 
    experiments of the $0 \nu \beta \beta$ decay.
\end{abstract}
\end{center}
\end{titlepage}
\renewcommand{\thefootnote}{\#\arabic{footnote}} 
\setcounter{footnote}{0}
%
\noindent {\it \textbf{Introduction\,:}}~
The fate of the lepton number is an interesting question 
of particle physics.  It is an accidental global symmetry
in the Standard Model (SM) and then 
physics beyond the SM can violate the symmetry.
One of the most interesting possibilities is the seesaw 
mechanism~\cite{Seesaw} for generating masses of active neutrinos.  
In this case right-handed neutrinos are introduced 
with Majorana masses which break the lepton number.
Thus, the non-zero masses of active neutrinos observed in oscillation 
experiments may lead to the violation if neutrinos are Majorana particles.

The lepton number violation induces striking processes which are
absent in the SM.  One famous example is the neutrinoless double
beta ($0 \nu \beta \beta$) decay: $(A,Z) \to (A, Z+2) + 2 \, e^-$,
which changes the lepton number by two units~\cite{Pas:2015eia}.  The
decay occurs if the neutrinos are massive Majorana particles and its
rate is proportional to the squared of the effective neutrino mass
$m_{\rm eff}$.  As for the $0 \nu \beta \beta$ decay of $^{136}$Xe,
the lower bound on the half-life is $T_{1/2} > 1.1 \times 10^{26}$~yr
at 90~\% C.L.~\cite{KamLAND-Zen:2016pfg} which leads to the upper bound on
the effective mass
$|m_{\rm eff} | < m_{\rm eff}^{\rm UB} = (61 - 161)$~meV, where the
uncertainties in Ref.~\cite{Faessler:2014kka} are taken into account.
The bound on the $0 \nu \beta \beta$ decay of $^{76}$Ge is
$T_{1/2} > 3.0 \times 10^{25}$~yr at 90~\%
C.L.~\cite{Agostini:2013mzu}.  In this case the bound becomes
$|m_{\rm eff} | < m_{\rm eff}^{\rm UB} = (213 - 308)$~meV.

The contribution to $m_{\rm eff}$ from active neutrinos 
$\nu_i$ ($i=1,2,3$), $m_{\rm eff}^\nu$, is given by
\begin{align}
  \label{eq:meffnu}
  m_{\rm eff}^\nu = \sum_{i=1,2,3} U_{e i}^2 \, m_i  \,,
\end{align}
where $m_i$ are masses of active neutrinos and
their ordering is $m_3 > m_2 > m_1$ in the normal hierarchy (NH) case
and $m_2 > m_1 > m_3$ in the inverted hierarchy (IH) case.
The mixing matrix of active neutrinos~\cite{PMNS}
 is denoted by
$U_{\alpha i}$ ($\alpha = e, \mu, \tau$), which is represented by
\begin{eqnarray}
  U = 
  \left( 
    \begin{array}{c c c}
      c_{12} c_{13} &
      s_{12} c_{13} &
      s_{13} e^{- i \delta} 
      \\
      - c_{23} s_{12} - s_{23} c_{12} s_{13} e^{i \delta} &
      c_{23} c_{12} - s_{23} s_{12} s_{13} e^{i \delta} &
      s_{23} c_{13} 
      \\
      s_{23} s_{12} - c_{23} c_{12} s_{13} e^{i \delta} &
      - s_{23} c_{12} - c_{23} s_{12} s_{13} e^{i \delta} &
      c_{23} c_{13}
    \end{array}
  \right)  
  \times
  \mbox{diag} 
  ( 1 \,,~ e^{i \eta} \,,~ e^{i \eta'}) \,,
\end{eqnarray}
where $s_{ij} = \sin \theta_{ij}$ and $c_{ij} = \cos \theta_{ij}$.
$\delta$ is the Dirac phase and $\eta$ and $\eta'$ are the Majorana
phases.  By using the central values of mixing angles $\theta_{ij}$ in
Ref.~\cite{Gonzalez-Garcia:2014bfa} and applying the cosmological bound
$\sum m_i < 0.23$~eV~\cite{Ade:2015xua}, the effective mass due to
active neutrinos is the range
$|m_{\rm eff}^\nu| = (1.49 - 72.0)~\mbox{meV}$ for the NH case or
$(18.5 - 82.4)$~meV for the IH case, respectively.  Thus, the
predicted range, especially in the IH case, will begin to be tested by
the near future experiments of the $0 \nu \beta \beta$ decay.

We revisit the $0 \nu \beta \beta$ decay in the neutrino Minimal
Standard Model ($\nu$MSM)~\cite{Asaka:2005an,Asaka:2005pn}, which is
the extension of the SM by three right-handed neutrinos $\nu_{R \, I}$ with masses below ${\cal O}(10^2)$~GeV.  The model
realizes the seesaw mechanism of neutrino masses and the baryogenesis
via oscillation of right-handed
neutrinos~\cite{Akhmedov:1998qx,Asaka:2005pn}, and offers a candidate
of dark matter (known as sterile neutrino dark matter~\cite{Dodelson:1993je,Adhikari:2016bei}).
The $0 \nu \beta \beta$ decay of the model
has been investigated
in Refs.~\cite{Bezrukov:2005mx,Asaka:2011pb,Asaka:2013jfa,Gorbunov:2014ypa}.  In this
model $m_{\rm eff}$ is induced not only by active neutrinos but also
by heavy neutral leptons (HNLs) associated with right-handed
neutrinos.  It has been shown that the contributions from HNLs can be
comparable to the one from active neutrinos and, importantly, it is a
destructive contribution due to the strong mass degeneracy of
HNLs~\cite{Asaka:2011pb}.

In this analysis we will show that HNLs can give a constructive
contribution to $m_{\rm eff}$ in a certain parameter region, which
should be contrast to previous results. 
This enhances the rate of the $0 \nu \beta \beta$ decay 
and hence it is a good target for the current and near future 
experiments. 

~

\noindent
{\it \textbf{$0 \nu \beta \beta$ decay in the $\nu$MSM\,:}}~
Let us consider the $\nu$MSM, in which 
three right-handed neutrinos $\nu_{R I}$ ($I=1,2,3$) are introduced
to the SM. Its Lagrangian is 
\begin{eqnarray}
  \label{eq:L_nuMSM}
  {\cal L}_{\nu{\rm MSM}} =
  {\cal L}_{\rm SM} +
  i \, \overline{\nu_R{}_I} \, \gamma^\mu \, \partial_\mu \, \nu_R{}_I
  -
  \Bigl(
  F_{\alpha I} \, \overline{L}_\alpha \, \Phi \, \nu_R{}_I
  + \frac{M_I}{2} \, \overline{\nu_R{}_I^c} \, \nu_R{}_I 
  + h.c.
  \Bigr)
\,,
\end{eqnarray}
where ${\cal L}_{\rm SM}$ is the SM Lagrangian.
$L_\alpha = (\nu_{L \alpha} \,,~ e_{L \alpha})^T$ and
$\Phi = (\phi^0\, ,~ \phi^-)^T$ are lepton and Higgs weak-doublets.
$M_I$ are Majorana masses of $\nu_{R I}$ (which are taken to be real
and positive) and $F_{\alpha I}$ are Yukawa coupling constants.  The
two types of neutrino masses,
$[M_D]_{\alpha I} = F_{\alpha I} \langle \Phi \rangle$ and $M_I$, are
assumed to be $|[M_D]_{\alpha I}| \ll M_I \lesssim {\cal O}(10^2)$
GeV.  The model realizes the seesaw mechanism and the mass eigenstates
in the neutrino sector are 
active neutrinos $\nu_i$ and heavy neutral leptons (HNLs) $N_I$.
The left-handed neutrinos are then written as
\begin{align}
  \nu_{L \alpha} = U_{\alpha i} \, \nu_i + \Theta_{\alpha I} \, N_I^c \,,
\end{align}
where the mixing matrix of HNLs is given by
$\Theta_{\alpha I} = [M_D]_{\alpha I} /M_I$.

The lightest HNL $N_1$ with keV scale mass is a candidate of dark
matter.  In order to avoid the constraints for dark matter its mixing
$\Theta_{\alpha 1}$ must be sufficiently small~\cite{Boyarsky:2009ix}.
Consequently, $N_1$ decouples from the seesaw mechanism and from the
baryogenesis.  The heavier ones $N_2$ and $N_3$ are then responsible
to these two mechanism.  It has been shown that they should be
quasi-degenerate $\Delta M = (M_3 - M_2)/2 \ll M_N = (M_3 + M_2 )/2$
and $M_N = {\cal O}(0.1)$--${\cal O}(10^2)$ GeV for the successful
scenario.  (See the analyses in
Refs.~\cite{Gorbunov:2007ak,Canetti:2010aw,Asaka:2013jfa,Blondel:2014bra}.)

The effective mass in the $\nu$MSM is given by
\begin{align}
  \label{eq:meffnuMSM}
  m_{\rm eff}^{\nu {\rm MSM}} = 
  m_{\rm eff}^\nu + m_{\rm eff}^N \,,
\end{align}
where the first term is the contribution from active neutrinos in
Eq.~(\ref{eq:meffnu}).  Since $N_1$ gives a suppressed contribution to
the masses of active neutrinos, the lightest one
is smaller than ${\cal O}(10^{-4})$ eV.  In this case, the
range of $m_{\rm eff}^\nu$ is limited as
$|m_{\rm eff}^\nu| = (1.49 - 3.66)~\mbox{meV}$ for the NH case or
$(18.5 - 47.9)$~meV for the IH case, respectively.  

On the other hand, the second term in Eq.~(\ref{eq:meffnuMSM}) denotes
the contributions from HNLs given by
\begin{align}
  m_{\rm eff}^N = \sum_{I=1,2,3} \Theta_{\alpha I}^2 \, 
  f_\beta (M_I) \, M_I \,,
\end{align}
where the $f_\beta (M_I)$ represents the suppression of the nuclear 
matrix element of the $0 \nu \beta \beta$ decay for 
$M_I \gg {\cal O}(0.1)$~GeV.  In this analysis, we follow
the results in Ref.~\cite{Faessler:2014kka} 
(see also Ref.~\cite{Barea:2015zfa})
and take the expression
\begin{align}
  f_\beta (M_I) = \frac{ \langle p^2 \rangle}
  {\langle p^2 \rangle + M_I^2 } \,,
\end{align}
with $\langle p^2 \rangle \simeq (200~\MeV)^2$.
As shown in Ref.~\cite{Bezrukov:2005mx} the contribution from $N_1$ is
much smaller than $m_{\rm eff}^\nu$ and those from $N_2$ and $N_3$
(see also Refs.~\cite{Asaka:2011pb,Merle:2013ibc}).
Thus, we shall neglect its contribution 
and set its Yukawa couplings $F_{\alpha 1} = 0$ from now on. 
In this case, the Yukawa couplings of 
$N_2$ and $N_3$ are parameterized as presented in Eq.~(2.8)
of Ref.~\cite{Asaka:2011pb}.  

The contributions from $N_2$ and $N_3$ is decomposed 
into two parts~\cite{Asaka:2011pb}
\begin{align}
  m_{\rm eff}^N = \sum_{I=2,3} \Theta_{\alpha I}^2 \, f_\beta (M_I) \, M_I 
  = \overline m_{\rm eff}^{N_{2,3}} + \delta m_{\rm eff}^{N_{2,3}} \,,
\end{align}
where
\begin{align}
  \label{eq:meffNbar}
  \overline m_{\rm eff}^{N_{2,3}} 
  &=
    f_{\beta}(M_N) \sum_{I=2,3} \Theta_{eI}^2 \, M_I 
    = - f_\beta (M_N) \, m_{\rm eff}^\nu 
    \,,
  \\
  \label{eq:dmeffN}
  \delta m_{\rm eff}^{N_{2,3}} 
  &=
    \sum_{I=2,3} [f_\beta(M_I) - f_\beta (M_N) ]
    \Theta_{eI}^2 \, M_I \,,
\end{align}
where we have used in Eq.~(\ref{eq:meffNbar}) the seesaw relation of
masses and mixings of active neutrinos and HNLs
\begin{align}
  \label{eq:SSR}
  0 = \sum_{i=1,2,3} U_{\alpha i}^2 \, m_i 
  + \sum_{I=1,2,3} \Theta_{\alpha I}^2 \, M_I \,.
\end{align}

In the previous analysis we have neglected $\delta m_{\rm eff}^{N_{2,3}}$
which vanishes for $\Delta M =0$.  This is because 
the baryogenesis via neutrino oscillations 
requires $\Delta M/M_N \ll 1$ in order to generate the sufficient 
amount of the baryon asymmetry of the universe (BAU). (See, however, the discussion below.)
In this case the effective mass in the $\nu$MSM is 
given by a simple expression~\cite{Asaka:2011pb}
\begin{align}
  \label{eq:meffnuMSM2}
  m_{\rm eff}^{\nu{\rm MSM}} =
  \left[ 1 - f_\beta(M_N) \right] \, m_{\rm eff}^\nu 
  =
  \frac{M_N^2}{ \langle p^2 \rangle + M_N^2 } \, m_{\rm eff}^\nu \,.
\end{align}
It is seen that the contributions from HNLs can be neglected when
$M_N^2 \gg \langle p^2 \rangle$~\cite{Bezrukov:2005mx}.  On the other
hand, when $M_N^2 \lesssim \langle p^2 \rangle$, $N_2$ and $N_3$ give
sizable, destructive contributions to $m_{\rm eff}$.  Especially, the
effective mass in the $\nu$MSM becomes zero as $M_N \to 0$.  This is a
general consequence of the seesaw relation~(\ref{eq:SSR}) when HNLs
are much lighter than $\langle p^2 \rangle^{1/2}$.

Let us summarize the results of the $0 \nu \beta \beta$ in the
$\nu$MSM so far: (1) HNLs $N_2$ and $N_3$ gives
a destructive contribution and then $m_{\rm eff}^{\nu{\rm MSM}}$
is given by Eq.~(\ref{eq:meffnuMSM2}), and hence
$|m_{\rm eff}^{\nu{\rm MSM}}| < | m_{\rm eff}^\nu |$.
(2) The upper bound on $|m_{\rm eff}^{\nu{\rm MSM}}|$
is the same as the upper bound on $|m_{\rm eff}^\nu|$ 
for the lightest active neutrino with ${\cal O}(10^{-4})$.
It is achieved for $M_N \gg \langle p^2 \rangle^{1/2}$.
(3) There is the lower bound on $|m_{\rm eff}^{\nu{\rm MSM}}|$
which is obtained by the smallest value of $|m_{\rm eff}^\nu|$ 
(for the lightest active neutrino with ${\cal O}(10^{-4})$)
multiplied by the factor $[1 - f_\beta (M_N)]$ with the smallest
value of $M_N$.  Notice that the lower bound on $M_N \simeq 150$~MeV
to avoid the cosmological constraint as well as the bounds from 
the direct search experiments 
(see the recent analysis~\cite{Asaka:2013jfa}).

~

\noindent {\it \textbf{Enhancement of~ $0 \nu \beta \beta$ decay in
    the $\nu$MSM\,:}}~
We are now at the position to discuss the main point of this analysis.
It will be shown that there is a certain parameter region in which the
effective mass $|m_{\rm eff}^{\nu{\rm MSM}}|$ can be larger than
$|m_{\rm eff}^{\nu}|$.  This is because of
$\delta m_{\rm eff}^{N_{2,3}}$ in Eq.~(\ref{eq:dmeffN}) which was
neglected in the previous analysis.  First, we note that
\begin{align}
  \delta m_{\rm eff}^{N_{2,3}}
  \simeq f'_\beta (M_N) \Delta M \,
  \left ( \Theta_{e 3}^2 \, M_3 - \Theta_{e 2}^2 \, M_2 \right)
  \simeq
  - \frac{2 \langle p^2 \rangle \, \langle \Phi^2 \rangle \, \Delta M}
  {(\langle p^2 \rangle + M_N^2 )^2}
  \left( F_{e3}^2 - F_{e2}^2 \right) \,,
\end{align}
where we have neglected the terms suppressed by
${\cal O}(\Delta M^2/M_N^2)$.  
Now we use the parameterization of the Yukawa coupling constants in 
Ref.~\cite{Asaka:2011pb}, and then obtain the expression
\begin{align}
  \delta m_{\rm eff}^{N_{2,3}}
  = 
  - \frac{\langle p^2 \rangle \, M_N^2 }
  {(\langle p^2 \rangle + M_N^2 )^2} 
  \frac{\Delta M}{M_N} \,  X_\omega^2 \, m_\ast \,,
\end{align}
where we have assumed $X_\omega \gg 1$.  Here the 
mass parameter $m_\ast$ is given by the mixing angles and 
phases as
\begin{align}
m_\ast =& 
e^{-2 i ({\rm Re \omega} + \delta)} 
\left( 
e^{i (\delta + \eta)} \sqrt{m_2} \sin \theta_{12} \cos \theta_{13} + i \xi \sqrt{m_3} \sin \theta_{13} \right)^2\,,
\end{align}
for the NH case and
\begin{align}
m_\ast =& 
e^{-2 i {\rm Re} \omega} \cos^2 \theta_{13} \left( \sqrt{m_1} \cos \theta_{12} + i \xi e^{i \eta} \sqrt{m_2} \sin \theta_{12} \right)^2\,,
\end{align}
for the IH case.  The mass parameter takes its maximal value
$|m_\ast| = 7.0$~meV if $\delta + \eta = \pi/2$ and $\xi = + 1$ (or
$\pi/3$ and $\xi = - 1)$ for the NH case, whereas $|m_\ast| = 92$~meV if
$\eta = 3 \pi/2$ and $\xi = + 1$ (or $\eta = \pi/2$ and $\xi = - 1$)
for the IH case.  Therefore, we find that
$\left| \delta m_{\rm eff}^{N_{2,3}} \right|$ can be significantly large,
when $M_N \simeq \langle p^2 \rangle^{1/2}$ and
$X_\omega \Delta M/M_N \gtrsim 1$, as
\begin{align}
  \left|\delta m_{\rm eff}^{N_{2,3}} \right|
  \simeq
  X_\omega^2 \,
  \left( \frac{\Delta M}{M_N} \right) \, |m_\ast| \,.
\end{align}
In order to discuss an impact of $\delta m_\text{eff}^{N_{2,3}}$, hereafter we restrict CP phases and $\xi$ as derived above. 
It should be noted that $\delta m_\text{eff}^{N_{2,3}}$ can always be a constructive contribution by choosing ${\rm Re} \omega$.
As we will demonstrate below, it can overcome $|m_{\rm eff}^\nu|$.
This point was missed in the previous analysis.

In the $\nu$MSM, the production mechanism of the lepton asymmetry is completely different from the ordinary leptogenesis. 
For the whole system the total lepton asymmetry is conserved at the early universe due to the smallness of the Majorana masses, 
and then the asymmetries are stored separately in left- and right-handed lepton sectors. 
Since the decay processes of right-handed neutrinos are irrelevant at the temperature above sphaleron freeze-out, 
the lepton asymmetry is produced through CP violating right-handed neutrino oscillation~\cite{Akhmedov:1998qx,Asaka:2005pn} which starts at the typical temperature $T_{\rm osc} = \left( M_0 \Delta M M_N / 6 \right)^{1/3}$ ($M_0 = 7.12 \times 10^{17}~{\rm GeV}$). 

In the baryogenesis via neutrino oscillations the upper bounds on
$X_\omega$ and $\Delta M$ can be obtained in order to generate the
observed baryon asymmetry~\cite{Canetti:2010aw}.  
When $X_\omega \gg 1$, the Yukawa coupling constants become larger which
leads to the washout of the produced asymmetry.  On the other
hand, when $\Delta M$ becomes larger, the oscillation of right-handed
neutrinos begins at earlier epoch, which again leads to the
suppression of the yield of baryon asymmetry.  In addition, when $M_N$
becomes smaller than $K$-meson mass, $N_{2}$ and $N_3$ can be produced
in $K$-meson decay and then receive the stringent constraints from
direct search experiments.  In this case, the mixing angles
$|\Theta_{\alpha 2}|$ and $|\Theta_{\alpha 3}|$ cannot be large and an
extremely large $X_\omega$ is disfavored. 
See, for example, the recent analysis in Ref.~\cite{Asaka:2014kia}.

\begin{figure}[t]
    \centerline{
    \includegraphics[clip, width=9.5cm]{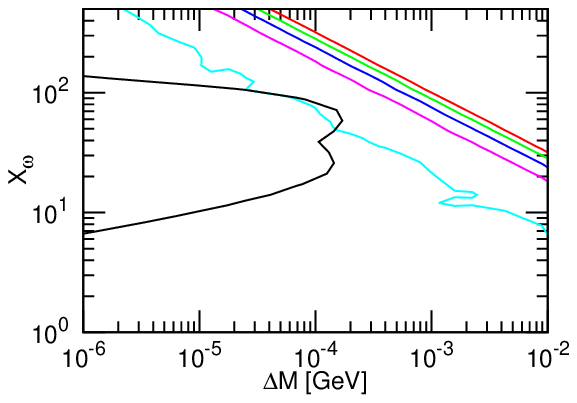}%
    \hspace{-1cm}\includegraphics[clip, width=9.5cm]{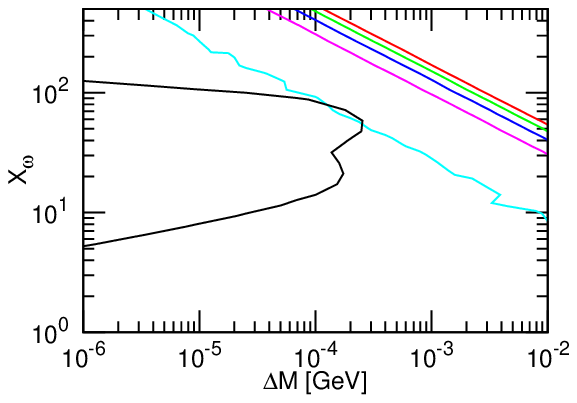}
    }
    \centerline{
      \hspace{3cm}(a)NH, $M_N=0.5$~GeV
      \hspace{5cm}
      (b)NH, $M_N=0.75$~GeV
    }
    \vspace{1cm}
    \begin{center}
        \includegraphics[clip, width=9.5cm]{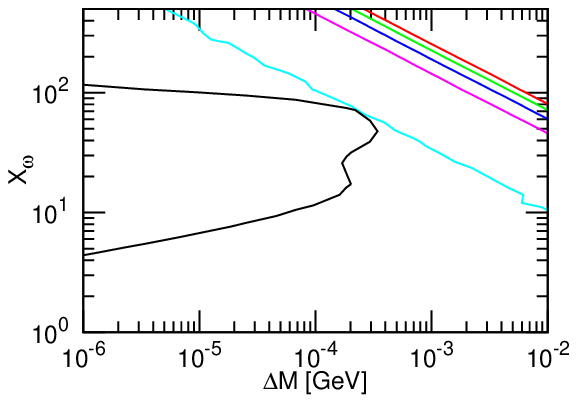}\\
        (c)NH, $M_N=1.0$~GeV
    \vspace{1cm}
    \caption{Contours of the maximal value of
      $|m_{\rm eff}^{\nu{\rm MSM}}|_{\rm max}$ in $\Delta M$-$X_\omega$ plane
      for the NH case.
      Here contour lines correspond to
      $|m_{\rm eff}^{\nu{\rm MSM}}|_{\rm max}\big/
      |m_{\rm eff}^{\nu}|_{\rm max}$ 
      = 1 (cyan line), 2 (magenta line), 3 (blue line), 4 (green line) 
      and 5 (red line).
      Here we take 
      $|m_\text{eff}|_\text{max}^\text{SM} = 3.66$~meV.
      In the region enclosed by the black line the BAU can be explained.}
    \label{fig:Max-NH}
  \end{center}
\end{figure}

\begin{figure}[t]
    \centerline{
    \includegraphics[clip, width=9.5cm]{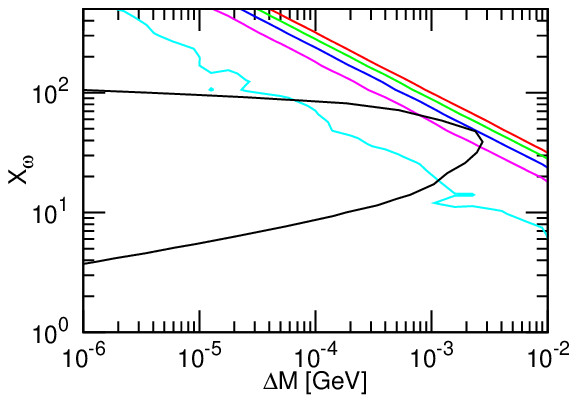}%
    \hspace{-1cm}\includegraphics[clip, width=9.5cm]{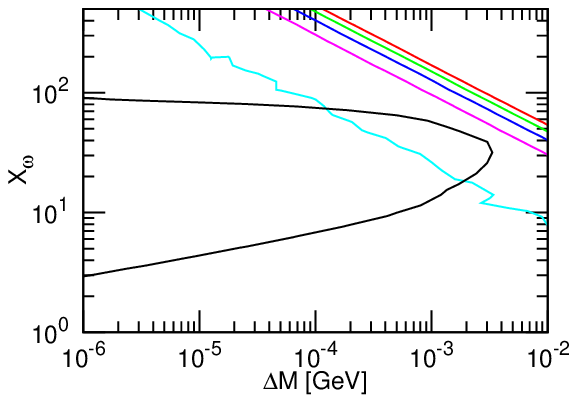}
    }
    \centerline{
      \hspace{3cm}(a)IH, $M_N=0.5$~GeV
      \hspace{5cm}
      (b)IH, $M_N=0.75$~GeV
    }
    \vspace{1cm}
    \begin{center}
        \includegraphics[clip, width=9.5cm]{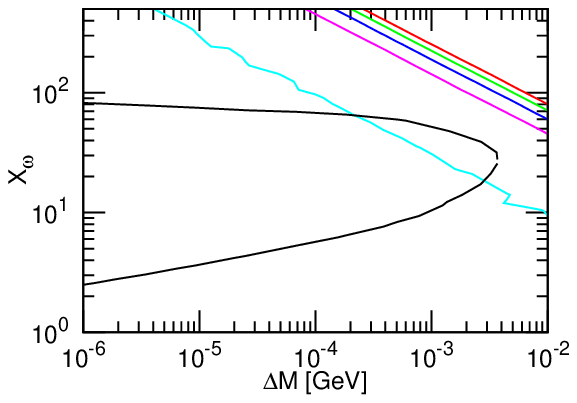}\\
        (c)IH, $M_N=1.0$~GeV
    \vspace{1cm}
    \caption{Contours of the maximal value of
      $|m_{\rm eff}^{\nu{\rm MSM}}|_{\rm max}$ in $\Delta M$-$X_\omega$ plane
      for the IH case.
      Here contour lines correspond to
      $|m_{\rm eff}^{\nu{\rm MSM}}|_{\rm max}\big/
      |m_{\rm eff}^{\nu}|_{\rm max}$ 
      = 1 (cyan line), 2 (magenta line), 3 (blue line), 4 (green line) 
      and 5 (red line).
      Here we take 
      $|m_\text{eff}|_\text{max}^\text{SM} = 47.9$~meV.
      In the region enclosed by the black line the BAU can be explained.}
    \label{fig:Max-IH}
  \end{center}
\end{figure}
Based on these arguments, we shall illustrate our idea in
Figs.~\ref{fig:Max-NH} and \ref{fig:Max-IH}.  Here we take
$M_N = 0.5$, 0.75, and 1~GeV for both NH and IH cases.  First, the
successful baryogenesis is possible in the region enclosed by the black line for each figure.

In this analysis, we use the kinetic equations of the case (i\makebox[-0.2mm][r]{}i) given in Ref.~\cite{Asaka:2011wq} for the estimation of the BAU. 
\begin{align}
\frac{d R_N}{dt} =& 
-i \left[ \langle H_N \rangle\,, R_N \right] 
- \frac{3 \langle \gamma_N^d \rangle}{2} \left\{ F^\dagger F\,, R_N - \bf{1} \right\} 
+ 2 \langle \gamma_N^d \rangle F^\dagger \left( A - \bf{1} \right) F 
- \frac{\langle \gamma_N^d \rangle}{2} \left\{ F^\dagger \left( A^{-1} - \bf{1} \right) F\,, R_N \right\}\,, \label{eq:KE_N}\\
\frac{d \mu_{\nu_\alpha}}{dt} =& 
- \frac{3 \gamma_\nu^d \left( T \right)}{2} \left[ F F^\dagger \right]_{\alpha \alpha} \tanh \mu_{\nu_\alpha} 
+ \frac{\gamma_\nu^d \left( T \right)}{2} \left[ F R_N F^\dagger - F^\ast R_{\bar{N}} F^T \right]_{\alpha \alpha} \frac{1}{\cosh \mu_{\nu_\alpha}} \notag\\
& 
+ \frac{\gamma_\nu^d \left( T \right)}{4} \left\{ \left[ F \left( R_N - \bf{1} \right) F^\dagger \right]_{\alpha \alpha} \left( 1 - \tanh \mu_{\nu_\alpha} \right) - \left[ F^\ast \left( R_{\bar{N}} - \bf{1} \right) F^T \right]_{\alpha \alpha} \left( 1 + \tanh \mu_{\nu_\alpha} \right) \right\}\,,
\end{align}
where $R_{N\,(\bar{N})}$ and $\mu_{\nu_\alpha}$ are the density matrix of right-handed (anti-) neutrino and the chemical potential of left-handed lepton, $A = \text{diag}(e^{\mu_{\nu_{e}}},e^{\mu_{\nu_{\mu}}},e^{\mu_{\nu_{\tau}}})$, 
$\gamma_\nu^d$ and $\gamma_N^d$ are destruction rates for left- and right-handed neutrinos, respectively.  
The equations for $\bar{N}$ can be obtained by the CP conjugation of Eq.~(\ref{eq:KE_N}).
$T$ is the temperature of the universe. 
(See the details in Ref.~\cite{Asaka:2011wq}.)
In Fig.~\ref{fig:Max-NH} and~\ref{fig:Max-IH}, the yield of the baryon asymmetry from these kinetic equations 
can be consistent with the observed value for the regions inside the black colored line.
The behavior of this line is interpreted as follows: 
The value of $X_\omega$ is bounded from both below and above as explained above.  
In addition, in the interesting parameter space where $\Delta M$ is relatively large,
the generated asymmetry at the sphaleron freeze-out temperature decreases 
as $\Delta M$ increases due to the suppression of the oscillation effect, 
and then $\Delta M$ is bounded from above. 

In these figures we also plot contours of the ratio
$\left| m_{\rm eff}^{\nu{\rm MSM}} \right|_{\rm max} \big/ \left| m_{\rm
      eff}^{\nu} \right|_{\rm max}$,
where the maximal value of the contribution from active neutrinos is
$\left| m_{\rm eff}^{\nu} \right|_{\rm max} = 3.66$ and 47.9~meV for
the NH and IH cases, respectively.  It is then found that the ratio
can be large as unity at most in the NH case within our setup.  
This means that 
the effective neutrino mass in the $\nu$MSM cannot exceed the active
neutrino one when the mass ordering is NH.
On the other hand, it should be noted that
the ratio can be large as three for the IH case, which means that
$\left| m_{\rm eff}^{\nu{\rm MSM}} \right|$ can be large as 
140~meV.  This should be contrast the results in 
the previous works.
Such a large value is realized when
$M_N \simeq$ 500 MeV, $\Delta M \simeq 10^{-3}$ GeV, and 
$X_\omega \simeq 50$.
Here the choice of these parameters gives the mixing angles of
heavy neutral leptons as~\cite{Asaka:2014kia}
\begin{align}
|\Theta|^2 = \sum_{\alpha = e, \mu, \tau} \sum_{I=2,3}
|\Theta_{\alpha I}|^2 =
\frac{\sum_{i=1,2,3} m_i}{2 M_N} X_\omega^2
\simeq 1.3 \times 10^{-7} .
\end{align}
Note that this parameter set gives rise to the large effective mass and the sufficient amount of the BAU at the same time. 
The whole of such possibilities can be derived be the numerical scan of the full parameter space of the model 
taking into account the experimental and cosmological constraints. 
This issue is, however, beyond the scope of the present analysis.

~

\noindent {\it \textbf{Summary and outlook\,:}}~
We have investigated the $0 \nu \beta \beta$ decay in the $\nu$MSM.
Especially, we have estimated the contribution
$\delta m_{\rm eff}^{{N_{2,3}}}$ proportional to the mass
difference $\Delta M$ between HNLs $N_2$ and $N_3$, 
which was not fully taken into account in the previous analysis.

It has been shown that the effective neutrino mass
$\left| m_{\rm eff}^{\nu{\rm MSM}} \right|$ can be large owing to this
effect  
if HNLs have a large mass difference and strong mixing and the common mass is close to the typical momentum of this process ($\langle p^2 \rangle^{1/2} \simeq 200 \ \text{MeV}$).  
Actually, we have shown that the maximal value of
$\left| m_{\rm eff}^{\nu{\rm MSM}} \right|$ is 140~meV when
$M_N \simeq$ 500 MeV, $\Delta M \simeq 10^{-3}$ GeV, and
$X_\omega \simeq 50$, if the mass ordering of active neutrinos is the
inverted one.  
Such a large value is comparable to the current upper
bound by using $0 \nu \beta \beta$ decay of $^{136}$Xe from KamLAND-Zen
experiment. 
It should be stressed that large $\left| m_{\rm eff}^{\nu{\rm MSM}} \right|$ is possible 
only if the CP phases including Majorana phase and the mixing angle of HNLs are aligned appropriately. 
Our analysis, therefore, indicates that these unexplored parameters of the $\nu$MSM start to be revealed 
by the $0 \nu \beta \beta$ decay  experiments.\footnote{While writing up this manuscript, one of the 
authors (S.E.) and M. Drewes presented numerical results of this issue in Ref.~\cite{Drewes:2016lqo} where 
the parameter scan including cosmological and experimental constraints is performed. In addition, after we submitted this work to arXiv,
we noticed that a paper (arXiv:1606.06719), which discusses the similar issue, appeared on the same day. 
These results are consistent with ours.
}

~

~

\noindent {\large \textbf{Acknowledgments}}\\
The work of T.A. was partially supported by JSPS KAKENHI
Grant Numbers 15H01031, 25400249, and 26105508.
T.A. thanks the Yukawa Institute for Theoretical Physics at Kyoto
University, where this work was completed during the YITP-S-16-01 on
"The 44th Hokuriku Spring School".
S.E. was supported by Swiss National Science Foundation 200020\_162927/1.



\end{document}